\documentclass{article}

\newif\ifEAfigs
\EAfigstrue

\newif\ifRuleLines
\RuleLinestrue

\newlength{\RuleWidth}
\newcommand{\BeginRule}{
\ifEAfigs \begin{figure}[htbp] 
\fi
\begin{center}
\ifRuleLines \rule{\RuleWidth}{.01in} \\ \fi
\begin{minipage}[t]{\RuleWidth}
\begin{em}
\begin{tabbing}
mmm\=mmm\=mmm\=mmm\=mmm\=mmm\=mmm\=mmm\=mmm\=mmm\=\kill
}

\newcommand{\EndRule}[2]{
\end{tabbing}
\end{em}
\end{minipage}
\ifRuleLines \rule{\RuleWidth}{.01in} \fi
\end{center}
\ifEAfigs  \caption{\label{#1} #2}
          \end{figure}  
\fi
}

\newcommand{\NoFigRules}{\EAfigsfalse}

\newcommand{\NoRuleLines}{\RuleLinesfalse}

\newcommand{\If}{{\bf if\ }}
\newcommand{\Then}{{\bf then\ }}
\newcommand{\Else}{{\bf else\ }}
\newcommand{\Elseif}{{\bf elseif\ }}
\newcommand{\Endif}{{\bf endif\ }}

\title{An Offline Partial Evaluator for Evolving Algebras}
\author{James K. Huggins%
\thanks{CSE Technical Report CSE-TR-229-95.
EECS Department, University of Michigan,
Ann Arbor, MI, 48109-2122, USA, huggins@umich.edu.
Partially supported by ONR grant 
N00014-91-J-1861 and NSF grant CCR-92-04742.}\\
}
\date{}

\begin{document}
\maketitle

\NoFigRules
\NoRuleLines

\begin{abstract}

We describe the architecture of an evolving algebra partial evaluator,
a program which specializes an evolving algebra with respect to
a portion of its input.  We discuss the particular analysis, 
specialization, and optimization techniques used and show an example
of its use.

\end{abstract}

\section{Introduction}

The Evolving Algebra Project was started by Yuri Gurevich as an attempt
to bridge the gap between formal models of computation and practical
specification methods.  The evolving algebras thesis is that any
algorithm can be modeled at its natural abstraction level by an
appropriate evolving algebra.  

Based upon this thesis, members of the evolving algebra community have
sought to develop a methodology based upon mathematics which would allow
algorithms to be modeled naturally; that is, described at their natural
abstraction levels.  The result is a simple methodology for describing
simple abstract machines which correspond to algorithms.  Plentiful
examples exist in the literature of evolving applied to different
types of algorithms (see \cite{biblio} for a current listing). 

The language of evolving algebras is extremely simple, consisting
chiefly of assignment and ``if-then'' statements.  Those familiar with
the partial evaluation literature will see similarites between evolving
algebras and Jones' flowchart language \cite{jgs}, although evolving
algebras (or \emph{ealgebras}) are massively parallel. 

In \cite{peval}, we introduced the idea of a partial evaluator
for ealgebras.  Ealgebras have often been used to
describe interpreters for programming languages; being able to
specialize these interpreters with respect to source programs would
allow one to automatically generate ealgebras for specific
programs.  Here we describe in greater detail the structure of an
offline partial evaluator for ealgebras.

\section{Sequential Evolving Algebras}

Sequential ealgebras are described fully in \cite{tutorial}; a more
formal description of ealgebras (including parallel and distributed
models) can be found in \cite{lipari}.  Here we recall the notions
behind basic sequential ealgebras.

Every ealgebra has a {\em vocabulary} (or {\em signature}); that is, a
finite collection of function names, each of a fixed arity.  Every
vocabulary contains certain obligatory names, including the nullary
function names {\em true}, {\em false}, {\em undef}, as well as the
names of the usual Boolean operations and the equality sign.  Function
names may be tagged as {\em static} and/or {\em relational}; the
significance of these tags will shortly become apparent.

A state $S$ of an ealgebra $\mathcal{D}$ with vocabulary $\Upsilon$ is
a non-empty set $|S|$, called the {\em superuniverse}, along with
interpretations of each function name in $\Upsilon$ over $S$.  The
interpretations of the nullary names {\em true}, {\em false}, and {\em
undef} are distinct in any $S$.  The interpretations of the Boolean
function names behave in the usual way over $\{\mathit{true},
\mathit{false}\}$ and take the value $\mathit{undef}$ otherwise.
Static function names have the same interpretation in any state $S$ of
a particular execution of an ealgebra.  The nullary name {\em undef}
is used to represent partial functions: a partial function $f$ takes
the value {\em undef} for argument tuples outside its intended domain.
Relations are represented as Boolean-valued functions.

Transition rules describe how states of an ealgebra change
over time.  An \emph{update instruction} is the simplest type of
transition rule and has the form
\BeginRule 
f($\bar{x}$) := $v$ 
\EndRule{}{} 
where $f$ is a non-static function name, $\bar{x}$ is a tuple of terms
of appropriate length, and $v$ is a term.  Executing such an
instruction has the expected result: if $\bar{a}$ and $a$ are the
values of $\bar{x}$ and $v$ in the current state, $f(\bar{a}) = a$
in the next state.

A \emph{rule block} is a transition rule and is simply a sequence
of transition rules.  To execute a rule block, execute each
of the rules in the sequence simultaneously.  Conflicts between
rules are not permitted. 

We also allow \emph{conditional instructions} of the form
\BeginRule 
\If $g_0$ \Then $R_0$\\
\Elseif $g_1$ \Then $R_1$\\ 
\vdots\\
\Elseif $g_n$ \Then $R_n$\\ 
\Endif\\ 
\EndRule{}{} 
where the $g_i$ are Boolean first-order terms and the $R_i$ are
transition rules.  (The phrase ``\Elseif true \Then $R_n$'' is usually
abbreviated as ``\Else $R_n$'').  To execute a transition rule of this
form in state $S$, evaluate the $g_i$ in state $S$; if any of the
$g_i$ evaluate to \emph{true}, execute transition rule $R_k$, where
$g_k$ is \emph{true} but $g_i$ is false for $i < k$.  If none of
the $g_i$ evaluate to \emph{true}, do nothing.

A \emph{program} for an ealgebra is a rule (usually, a rule block).
A {\em run} of an ealgebra from an initial state $S_0$ is a sequence of
states $S_0, S_1, \ldots$ where each $S_{i+1}$ is obtained from $S_i$ by
executing the program of the algebra in state $S_i$. 

\subsection{Pre-Processor}

Our eventual goal is to be able to execute as many rules of the
ealgebra program at specialization time as possible.  It will make our
specializer simpler if the input program were structured in a
restricted manner.

The pre-processor performs three types of transformations.
First, all transition rules are examined to see if any rule blocks
contain rule blocks as members.  In any such rule, the members
of the inner rule block are ``promoted''; that is, the inner
block is removed and its constituent members are added to the
outer block. 

Since all transition rules fire
simultaneously, the following transition rules are equivalent:
\BeginRule
\If $g_0$ \Then $R_0$   \hspace{3cm} \=\If $g_0$ \Then $R_0$ \Else\\
\Elseif $g_1$ \Then $R_1$     \>\ \ \ \If $g_1$ \Then $R_1$ \Else\\
\vdots                        \>\ \ \ \ \vdots\\
\Elseif $g_k$ \Then $R_k$     \>\ \ \ \ \ \ \ \If $g_k$ \Then $R_k$ \Endif\\
\Endif                        \>\ \ \ \ \vdots\\
                              \>\ \ \ \Endif\\
                              \>\Endif\\

\EndRule{}{}

The pre-processor continues by applying this transformation to
all transition rules.  The resulting program contains only simple
``\textbf{if}-\textbf{then}-\textbf{else}'' statements.

The following transition rules are also equivalent:
\BeginRule
$R_0$\\
\If \=guard \Then  \hspace{3cm}\=\If \=guard \Then\\   
	\>$R_1$			\>      \>$R_0$\\
	\>			\>   	\>$R_1$\\
	\>			 \>      \>$R_3$\\
( \Else   \>                      \>( \Else\\
        \>$R_2$ )		\>      \>$R_0$\\
\Endif  \>	                \>      \>$R_2$\\  
$R_3$   \>                      \>      \>$R_3$ )\\  
        \> 			\>\Endif\\          
\EndRule{}{}
Our pre-processor repeatedly applies this transformation to the input
program as long as possible (allowing for $R_0$ or $R_3$ to be empty).
This has the effect of pushing the update instructions as far into the
nesting of \textbf{if} statements as possible.  This transformation
usually increases the size of the program, possibly exponentially;
we restrict our attention to programs in which this transformation
is feasible.

At this point, every transition rule has one of the following forms:

\begin{itemize}
\item An update instruction
\item An \emph{update block}, that is, a sequence of
update instructions
\item A guarded rule 
\BeginRule
\If guard \Then $R_1$ \Else $R_2$ \Endif
\EndRule{}{}
where $R_1$ and $R_2$ are simplified transition rules.
\end{itemize}
Note that a sequence of guarded rules is not a transition rule under
this definition.  For the rest of this paper we will assume that
transition rules have this form.

Such a program can be represented as a binary tree, where the leaves
are update blocks and the internal nodes are Boolean guards.
Thus, to execute a simplified transition rule, one traverses the tree,
evaluating the Boolean guards encountered at each node, eventually
reaching a sequence of updates which should be executed.

The benefits of this restructuring will become apparent when
we discuss the specializer in section~\ref{spec}.

\subsection{Binding-Time Analyzer}

The binding-time analyzer partitions the functions%
\footnote{More precisely, we partition the function names, since the
actual functions themselves are not known until specialization time.}
of the evolving algebra into two sets: \emph{positive} functions to be
pre-computed at specialization time, and \emph{negative} functions
which cannot or should not be pre-computed.  (The terms \emph{static}
and \emph{dynamic} are often used in the literature to describe this
distinction; however, these terms have different meanings for evolving
algebras.)

Each program has a set of \emph{input functions}; that is, functions
whose values are supplied by the user when the program is run.  The
user supplies the binding-time analyzer with a partition of the input
functions into two sets: \emph{input positive} functions whose initial
values will be supplied by the user at specialization time, and
\emph{input negative} functions whose initial values may not be
available at specialization time.

We say that function $f$ is \emph{directly dependent} on function $g$
in program $\mathcal{P}$ if an update $f(\bar(t)) := t_0$ appears
in $\mathcal{P}$ such that $g$ appears in $\bar(t)$ or $t_0$.  
A function $f$ is \emph{dependent} on function $g$ in program 
$\mathcal{P}$ if there exist functions $h_1, h_2, \ldots, h_k$ 
such that $f = h_1$, $g = h_k$, and $h_i$ is directly dependent
on $h_{i+1}$ for $1 \leq 1 < k$.   

The analyzer begins by marking as (finally) negative any input
negative function.  Next, any function dependent upon another negative
function within an update is marked as negative.  

At this stage, we know that no non-negative function depends upon any
input negative function.  Thus, every non-negative function depends
only upon functions whose values are known in the initial state, and
thus its values could be computed at specialization time.  However,
such a function might take infinitely many values during execution of
the program, and could lead the specializer into an infinite
computation if it were marked positive.

The analyzer continues by marking as (finally) positive any input
positive static function (\emph{i.e.}, a function which is not updated
within the program and is known in the initial state); such functions
never change during execution and are safe for specialization.  Next
the analyzer marks positive any function which depends only on other
positive functions; by a simple inductive argument, such positive
functions take only finitely many values and are thus safe for
specialization.

At this stage, functions which are neither positive or negative are
either self-dependent (that is, dependent upon itself) or dependent on
other self-dependent functions.  Without further information about the
functions in question, there is little that can be done to determine
whether these functions can be safely classified as positive.

\cite{jgs} gives one method for resolving this question for
self-dependent functions based upon well-founded partially ordered
domains.  We have implemented a version of this algorithm.  Other
resolution schemes are certainly possible and are contemplated for
future versions of the analyzer.

Any remaining unclassified functions are classified as negative.

\subsection{Specializer}
\label{spec}

Our binding-time analysis has identified a set of positive functions
to be pre-computed by the specializer.  The specializer will generate
code in which references to these positive functions will be replaced
by their known values.

The program produced by the specializer uses an additional function
$K$ (an allusion to ``known'').  $K$ will take on values which are
\emph{reduced states}: states of the ealgebra in which all
negative functions have been removed.

For notational clarity, we define a \emph{positive expression} to be
an expression composed solely of positive functions.  A \emph{negative
expression} is any expression which is not positive.

We denote the specialization of a transition rule $R$ with respect
to a particular reduced state $\kappa$ by $R_\kappa$.  For
a given $R$, $R_\kappa$ is defined as follows:


\begin{itemize}
\item Specializing a conditional instruction whose guard $g$ is
a negative expression and whose \textbf{then} and \textbf{else}
branches are $T$ and $E$ yields the following rule:
\BeginRule
\If $g_\kappa$ \Then $T_\kappa$ \Else $E_\kappa$ \Endif\\
\EndRule{}{}

\item Specializing a conditional instruction whose guard $g$ is a
positive expression and whose \textbf{then} and \textbf{else} branches
are $T$ and $E$ yields $T_\kappa$ if $g$ evaluates to $\mathit{true}$
in $\kappa$ and $E_\kappa$ otherwise.

\item Specializing an update block with updates 
$p_1, \ldots, p_k$ to positive functions and updates $n_1, \ldots,
n_l$ to negative functions yields the block composed of the
members of $P_\kappa$ and $N_\kappa$, where $P = p_1, \ldots, p_k$ and
$N = n_1, \ldots, n_l$.

\item Specializing an update block composed of updates $n_1, \ldots,
n_l$ to negative functions yields updates $n_{1\kappa}, \ldots,
n_{l\kappa}$, where each $n_{i\kappa}$ is formed by replacing
all positive expressions within $n_i$ by their values in $\kappa$.

\item Specializing an update block composed of updates $p_1, \ldots,
p_k$ to positive functions yields an update $K := \kappa'$, where
$\kappa'$ is the reduced state generated by applying $p_1, \ldots,
p_k$ to $\kappa$.
\end{itemize}

Here we can see why we transformed our program 
during the pre-processing phase.  Updates to $K$ are generated
from sequences of updates to positive functions.  Thus, in order to
specialize a transition rule with respect to a particular $K$-value,
we need to find all updates to positive functions across the program
which fire under similar conditions.  After pre-processing, all such
updates appear at the lowest possible nesting level.

A \emph{K-rule} is a transition rule of the form
\BeginRule 
\If K = $\kappa$ \Then $R_\kappa$ \Endif 
\EndRule{}{}
where $R_\kappa$ is the original program specialized with respect 
to $\kappa$.  

The specializer begins with the values of the positive functions
supplied by the user for the initial state.  It creates a $\kappa$ for
this state and generates a K-rule with respect to $\kappa$.  It then
continues to generate K-rules for different reduced states $\kappa$
until every $\kappa$ which appears within a K-rule has been used to
generate a K-rule.  The output of the specializer is the collection of
K-rules generated.

Since each positive function can take only finitely many values, only
finitely many values for $K$ can be generated by this process.  Thus,
this process will eventually terminate.  $K$ may possibly take 
on an infeasible (but finite) number of values; we restrict our
attention to programs in which $K$ can be feasibly computed.

\subsection{Optimizer}

The optimizer performs several local optimizations on the program
produced by the specializer in order to produce a shorter program.
These optimizes have been shown to be effective in practice in producing
shorter, more efficient code.  They are not comprehensive; other
optimization techniques may provide further enchancements. 

\subsubsection{Live Term Analysis}

Often our specializer produces K-rules which make use of expressions
that are used only briefly in the lifetime of the program.  For example,
consider the following K-rules:
\BeginRule
\If K=$k_1$ \Then b := c, K := $k_2$, \ldots \Endif\\
\If K=$k_2$ \Then a := b, K := $k_3$, \ldots \Endif\\
\EndRule{}{}
where $c$ occurs nowhere else in the rules shown above.
If no other assignment ``K := $k_2$'' appears anywhere in the program,
we can replace these two rules by the following:
\BeginRule
\If K=$k_1$ \Then K := $k_2$, \ldots \Endif\\
\If K=$k_2$ \Then a := c, K := $k_3$, \ldots \Endif\\
\EndRule{}{}
To make optimizations such as the above more systematically, we
perform a live code analysis (similar to one described in \cite{asu}).
Any instance of a term such as \emph{b} above which serves strictly as
an alias for another term is replaced by the term it aliases.

\subsubsection{Compatible Rule Analysis}

Consider the following two K-rules:
\BeginRule
\If K=$k_1$ \Then a := b, K := $k_2$ \Endif\\
\If K=$k_2$ \Then c := d, K := $k_3$ \Endif\\
\EndRule{}{}
Since the second K-rule above does not use function \emph{a}, the two
blocks of rules are compatible.  Thus, if no other assignment ``K :=
$k_2$'' appears in the program, we can replace these two rules by the
following:
\BeginRule
\If K=$k_1$ \Then a := b, c := d, K := $k_3$ \Endif\\
\EndRule{}
The optimizer performs an analysis of this form, combining rules
which are not inter-dependent.  

\subsubsection{Analysis of Unnecesssary IFs}

Occasionally, the above rule optimizations result in ``\textbf{if}''
statements being generated of the form
\BeginRule
\If guard \Then $R$ \Else $R$ \Endif\\
\EndRule{}{}
which can be replaced simply by the rule
\BeginRule
$R$
\EndRule{}{}
without any change to the meaning of the program.  The analyzer
checks all transition rules in the optimized program and removes
any unnecessary guards from within ``\textbf{if}'' statements.

\subsubsection{Speciality Optimizations}

Finally, a few speciality optimizations (e.g. transforming
\emph{Car(Cons(A,B))} to \emph{A}) are performed.  Of course, these
are highly dependent upon the particular functions in use in the program.

\section{Evaluating the Evaluator}

The partial evaluator described here has been implemented in C.
It has performed well on several small-scale tests.

Consider, for example, the following fragment of code
written in C:

\begin{center}
{\tt {void strcpy (char *s, char *t) \verb+{+ 
                while (*s++ = *t++) ; \verb+}+ } }
\end{center}

This function copies a string from the memory location indicated by 
$t$ to the memory location indicated by $s$.  It is admittedly
cryptic.

In \cite{c}, we presented an ealgebra interpreter for the C
programming language.  As a test, we ran our partial evaluator on our
algebra for C, specializing it with respect to {\tt strcpy()}.  
We reported in \cite{peval} the results of this test.  Improvements
to the optimizer have allowed us to generate the following result
directly (with most of the terms renamed for clarity):
\BeginRule
\If K = ``init'' \Then\\
    \>To := {\tt s}, From := {\tt t}, K := ``loop''\\
\Elseif K = ``loop'' \Then\\
    \>\If Memory(From) = 0 \Then\\
    \>   \>To := To + 1, From := From + 1\\
    \>   \>Memory(To) := Memory(From), K := ``loop''\\
    \>\Else\\
    \>   \>To := To + 1, From := From + 1\\
    \>   \>Memory(To) := Memory(From), K := ``done''\\
\Endif
\EndRule{}{}

In some sense this program might be called ``optimal'', in that every
update performed by this evolving algebra corresponds to an action
which must be taken by the original {\tt strcpy} code.  It could be
expressed more concisely (notice that the update ``\emph{To := To +
1}'' appears within both branches of an \textbf{if} and could thus be
moved outside of the \textbf{if}); which presentation is preferable 
depends on the intended use of the specialized program.

We have had some success applying the self-interpreter test
\cite{jgs}.  A self-interpreter for evolving algebras is an evolving
algebra which takes another evolving algebra as input and executes it.
Partially evaluating a self-interpreter with respect to a target
program should yield the target program (or something quite similar).

We have successfully applied our partial evaluator to the
self-interpreter, given the self-interpreter as input.  The output is
extremely similar to that of the self-interpreter.  It does contain
many updates which serve as aliases for constant terms which are
generated through the optimization process; however, the optimizer is
currently unable to detect these useless updates and remove them.

Future work on a partial evaluator for evolving algebras may focus on
the development of an on-line version which may produce even better
output that the off-line version, or the production of a
self-applicable partial evaluator.

\paragraph*{Acknowledgements.}  Our partial evaluator is based 
upon an evolving algebras interpreter developed with Ben Harrison and
Yuri Gurevich.  Yuri Gurevich also supervised this work, providing
helpful guidance with difficult problems as well as critical
commentary on early drafts of this paper.

\end{document}